\documentclass[onecolumn,showpacs]{revtex4}

\usepackage{graphicx}
\usepackage{dcolumn}
\usepackage{bm}

\input epsf

\begin{document}

\title{\Large Role of Initial Data in Higher Dimensional Quasi-Spherical Gravitational Collapse}

\author{\bf Ujjal Debnath}
\email{ujjaldebnath@yahoo.com}
\author{\bf Subenoy Chakraborty}
\email{subenoyc@yahoo.co.in}

 \affiliation{Department of
Mathematics, Jadavpur University, Calcutta-32, India.}

\date{\today}

\begin{abstract}
We study the gravitational collapse in ($n+2$)-D quasi-spherical
Szekeres space-time (which possess no killing vectors) with dust
as the matter distribution. Instead of choosing the radial
coordinate `$r$' as the initial value for the scale factor $R$, we
consider a power function of $r$ as the initial scale for the
radius $R$. We examine the influence of initial data on the
formation of singularity in gravitational collapse.
\end{abstract}

\pacs{0420D, 0420J, 0470B}

\maketitle

Over the last two decades gravitational collapse in spherical
space-time (TBL model) has been studied extensively with dust as
the matter content [1-10]. A general conclusion from these studies
is that a central curvature singularity may be naked but its local
or global visibility depends on the choice of initial data. If
the regular initial density profile falls off rapidly having a
maximum value at the centre then it is not possible to have naked
singularity above five dimensional space-time [8-10].\\

Also from the recent past, attention has been given to study
non-spherical collapse [11-20]. Most of these studies deal
collapse numerically [12-15] with a few analytical works [16] (for
quasi-spherical gravitational collapse, see ref. [17-20]). These
are mainly concerned with special shape of the gravitating body.
The present work examines the role of initial data in the
formation of
gravitational collapse in ($n+2$)-D Szekeres space-time.\\

The metric ansatz for ($n+2$)D Szekeres space-time is [21, 22]

\begin{equation}
ds^{2}=dt^{2}-e^{2\alpha}dr^{2}-e^{2\beta}\sum^{n}_{i=1}dx_{i}^{2}
\end{equation}

where $\alpha$ and $\beta$ are functions of all the ($n+2$)
space-time co-ordinates. But if we assume that $\beta'\ne 0$ then
for inhomogeneous dust model the solution is [22]

\begin{equation}
e^{\beta}=R(t,r)~e^{\nu(r,x_{1},...,x_{n})}
\end{equation}

\begin{equation}
e^{\alpha}=\frac{R'+R~\nu'}{\sqrt{1+f(r)}}
\end{equation}

\begin{equation}
e^{-\nu}=A(r)\sum_{i=1}^{n}x_{i}^{2}+\sum_{i=1}^{n}B_{i}(r)x_{i}+C(r)
\end{equation}

\begin{equation}
\dot{R}^{2}=f(r)+\frac{F(r)}{R^{n-1}}
\end{equation}

\begin{equation}
\rho(t,r,x_{1},...,x_{n})=\frac{n}{2}~\frac{F'+(n+1)F\nu'}{R^{n}(R'+R\nu')}
\end{equation}

\begin{equation}
\sum_{i=1}^{n}B_{i}^{2}-4AC=-1
\end{equation}

where $A(r),~ B_{i}(r)$ and $C(r)$ are arbitrary functions of $r$
satisfying (7) and also in the expression (5), $f(r)$ and $F(r)$
are arbitrary functions of $r$ alone.\\

A shell focusing singularity on a shell of dust occurs when it
collapses at or starts expanding from the centre of matter
distribution. We shall consider only the central shell focusing
singularity (i.e., $R=0$ or $\beta=-\infty$) for marginally bound
case only (i.e., $f(r)=0$). Suppose $t=t_{i}$ be the initial
hypersurface from which the collapse develops. For initial data
we assume that $R(t_{i},r)$ is a monotone increasing function of
$r$. So without any loss of generality, it is possible to make an
arbitrary relabeling of the dust shells by $r\rightarrow g(r)$
such that we can choose
\begin{equation}
R(t_{i},r)=R_{0}r^{k},~~~~~(R_{0}>0,~~ k~~ are~ constants)
\end{equation}

Hence solving equation (5) using the initial condition (8) we get
\begin{equation}
R=\left[R_{0}^{\frac{n+1}{2}}r^{\frac{(n+1)k}{2}}-\frac{n+1}{2}\sqrt{F(r)}~(t-t_{i})\right]^{\frac{2}{n+1}}
\end{equation}

The regularity of the metric co-efficients on the initial
hypersurface restricts $k\ge 1$. Further for the regularity of
the initial density profile
\begin{equation}
\rho_{i}(r,x_{1},...,x_{n})=\rho(t_{i},r,x_{1},...,x_{n})=
\frac{n}{2}~\frac{F'+(n+1)F\nu'}{R_{0}^{n+1}r^{(n+1)k-1}(k+r\nu')}
\end{equation}

we can write the following series expansion for $\rho_{i}(r)$,
$F(r)$ and $\nu'(r)$ near $r=0$ as [22]

\begin{equation}
\rho_{i}(r)=\sum_{j=0}^{\infty}\rho_{j}~r^{j},
\end{equation}

\begin{equation}
F(r)=\sum_{j=0}^{\infty}F_{j}~r^{(n+1)k+j}~~
\end{equation}
and
\begin{equation}
\nu'(r)=\sum_{j=-1}^{\infty}\nu_{j}~r^{j},~~~(\nu_{_{-1}}+k\ge 0)
\end{equation}

Also using equations (11)-(13) in (10) we have the following
relations among the different co-efficients
\begin{eqnarray*}
\rho_{0}=\frac{n(n+1)}{2}F_{0}R_{0}^{-(n+1)},~~\rho_{1}=\frac{n}{2}\left(n+1+\frac{1}{k+
\nu_{_{-1}}}\right)F_{1}R_{0}^{-(n+1)},
\end{eqnarray*}
\vspace{-5mm}

\begin{eqnarray*}
\rho_{2}=\frac{n}{2}\left[\left(n+1+\frac{2}{k+
\nu_{_{-1}}}\right)F_{2}-\frac{F_{1}\nu_{_{0}}}{(k+\nu_{_{-1}})^{2}}\right]R_{0}^{-(n+1)},
\end{eqnarray*}
\vspace{-5mm}

\begin{equation}
\rho_{3}=\frac{n}{2}\left[\left(n+1+\frac{3}{k+
\nu_{_{-1}}}\right)F_{3}-\frac{2F_{2}\nu_{_{0}}}{(k+\nu_{_{-1}})^{2}}-
\frac{(k+\nu_{_{-1}})\nu_{_{1}}-\nu_{_{0}}^{2}}{(k+\nu_{_{-1}})^{3}}F_{1}\right]R_{0}^{-(n+1)},
\end{equation}
$$...~~...~~...~~...~~...~~...~~...~~...,$$

$$OR$$

\begin{eqnarray*}
\rho_{0}=\frac{n}{2}\left[\frac{F_{1}}{\nu_{0}}+(n+1)F_{0}\right]R_{0}^{-(n+1)},~~
\rho_{1}=\frac{n}{2}\left[\frac{2F_{2}}{\nu_{0}}+\left\{(n+1)-
\frac{\nu_{1}}{\nu_{0}^{2}}\right\}F_{1}\right]R_{0}^{-(n+1)},
\end{eqnarray*}
\vspace{-5mm}

\begin{eqnarray*}
\rho_{2}=\frac{n}{2}\left[\frac{3F_{3}}{\nu_{0}}+\left\{(n+1)-
\frac{2\nu_{1}}{\nu_{0}^{2}}\right\}F_{2}+\left(\frac{\nu_{1}^{2}}{\nu_{0}^{3}}-
\frac{\nu_{2}}{\nu_{0}^{2}}\right)F_{1}\right]R_{0}^{-(n+1)},
\end{eqnarray*}
\vspace{-5mm}

\begin{equation}
\rho_{3}=\frac{n}{2}\left[\frac{4F_{4}}{\nu_{0}}+\left\{(n+1)-
\frac{3\nu_{1}}{\nu_{0}^{2}}\right\}F_{3}+2\left(\frac{\nu_{1}^{2}}{\nu_{0}^{3}}-
\frac{\nu_{2}}{\nu_{0}^{2}}\right)F_{2}+\left(\frac{2\nu_{1}\nu_{2}}{\nu_{0}^{3}}-
\frac{\nu_{3}}{\nu_{0}^{2}}-\frac{\nu_{1}^{3}}{\nu_{0}^{4}}\right)F_{1}\right]R_{0}^{-(n+1)},
\end{equation}
$$...~~...~~...~~...~~...~~...~~...~~...,$$

according as $\nu_{_{-1}}>-k$ ~or~ $\nu_{_{-1}}=-k.$\\

The singularity curve $t=t_{s}(r)$ for the shell focusing
singularity is characterized by

\begin{equation}
R(t_{s}(r),r)=0
\end{equation}

So the time for central shell focusing singularity is given by

\begin{equation}
t_{0}=t_{s}(0)=t_{i}+\frac{2R_{0}^{\frac{n+1}{2}}}{(n+1)\sqrt{F_{0}}}
\end{equation}

Further, if $t_{ah}(r)$ is the instant for the formation of
apparent horizon then we have

\begin{equation}
R^{n-1}(t_{ah}(r),r)=F(r)
\end{equation}

which gives

\begin{eqnarray*}
t_{ah}(r)-t_{0}=-\frac{R_{0}^{\frac{n+1}{2}}}{(n+1)F_{0}^{3/2}}\left[F_{1}~r+\left(F_{2}-\frac{3F_{1}^{2}}{4F_{0}}
\right)~r^{2}+... ...\right]
\end{eqnarray*}
\vspace{-5mm}

\begin{equation}
-\frac{2}{n+1}F_{0}^{\frac{1}{n-1}}\left[r^{\frac{(n+1)k}{n-1}}+\frac{1}{n-1}\frac{F_{1}}{F_{0}}~r^{\frac{(n+1)k}
{n-1}+1}+... ...\right]
\end{equation}

The above expression shows the time difference between the
formation of trapped surface at a distance $r$ and the time of
singularity formation at $r=0$ (central singularity). Hence the
necessary condition that an observer at a distance $r$ will
observe the central singularity (at least locally) is
$t_{ah}(r)>t_{0}$ (for details see Ref. [6$-$9, 24]).\\

From physical consideration it is reasonable to assume that the
initial density $\rho_{i}(r)$ is maximum at the centre $r=0$.
This implies that the first non-vanishing term after $\rho_{0}$
in the series expansion (see eq. (11)) for $\rho_{i}(r)$ should
be negative. Further, one may assume that $\rho'_{i}(r)$ should
vanish at $r=0$ but is negative in the neighbouring region or
more generally, it may be assumed that
$\rho'_{i}(r)=\rho''_{i}(r)=...=\rho^{s-1}_{i}(r)=0$ and
$\rho^{s}_{i}(r)<0$ ($s\ge 2$). Now using the relations (14) or
(15) among the co-efficients we have\\

I.~~ when $\nu_{_{-1}}>-k$:\\

Here $\rho_{1}<0$ implies $F_{1}<0$. Also in general,
$\rho_{1}=\rho_{2}=...=\rho_{s-1}=0$ and $\rho_{s}<0$ implies
$F_{1}=F_{2}=...=F_{s-1}=0$ and $F_{s}<0$ with $s\ge 2$.\\

II.~~ when $\nu_{_{-1}}=-k$:\\

Here $\rho_{1}<0$ does not imply $F_{1}<0$. Also more generally,
$\rho_{1}=\rho_{2}=...=\rho_{s-1}=0$ and $\rho_{s}<0$ may have
one possible solution as  $F_{1}=F_{2}=...=F_{s}=0$ and $F_{s+1}<0$
with $s\ge 2$.\\

In particular, if we assume the initial density to have a maximum
value at $r=0$ and falls off rapidly near $r=0$ then we have
$\rho_{1}=0, \rho_{2}<0$ near $r=0$. Then in case I, we have
$F_{1}=0, F_{2}<0$ while in case II, we may take $F_{1}=F_{2}=0,
F_{3}<0$ near $r=0$.\\

Therefore in the present problem we have the following
possibilities for naked singularity:\\

($a$)~~ If $F_{1}<0$ then naked singularity may appear in any
dimension ($\ge 4$) for $\nu_{_{-1}}\ge -k$ and $k\ge 1$.\\

($b$)~~ In general if we choose $F_{1}=0, F_{2}=0,...,F_{i-1}=0$
and $F_{i}<0$ ($i\ge 2$) then for formation of naked singularity,
`$n$' is restricted by the inequality

\begin{equation}
2\le n \le \left[\frac{i+k}{i-k}\right]
\end{equation}

with~~~~~ $max(1, \frac{i}{3})\le k<i$, ~~ $\nu_{_{-1}}\ge -k$,
~~$k\ge 1$.\\

Here $[x]$ stands for the greatest integer in $x$. From the
inequality (20) we note that if `$k$' is very close to `$i$' (but
less than $i$ ) then `$n$' can take larger values than 2 i.e.,
naked singularity may appear in much larger dimension compare to
the usual four dimension. The following table shows some possible
values of $n$ for different values of $i$ and $k$ (with $k<i$)
from the inequality (20).\\

\[
\text {TABLE:~~ Some possible values of $n$ for different values
of $i$ and $k$.} \]
\[
\begin{tabular}{|l|l|r|r|r|}\hline
\hline $~i\rightarrow$ & \multicolumn{1}{c|}{2} &
\multicolumn{1}{c|}{ 3} & \multicolumn{1}{c|}{ 4} &
\multicolumn{1}{c|}{5} \\$k\downarrow$ &  &  &  &
\\\hline
&  &  &  &  \\
$\frac{3}{2}$ & $n=2,3,4,5,6,7$ & $n=2,3~~~~$ & $n=2$~~~~~~ & $-$~~~~ \\
&  &  &  &  \\
\hline
&  &  &  &  \\
$\frac{21}{11}$ & ~~~~$2\le n\le 41$ & $n=2,3,4$~~ & $n=2$~~~~~~ & $-$~~~~ \\
&  &  &  &  \\
\hline
&  &  &  &  \\
$\frac{35}{12}$ & ~~~~~~~~~$-$ & $2\le n\le 71$ & $n=2,3,4,5,6$ & $n=2,3$ \\
&  &  &  &  \\
\hline\hline
\end{tabular}%
\]%
\newline

For $\nu_{_{-1}}>-k$ we have from equation (14) $F_{1}=0,
F_{2}<0$ and hence for naked singularity

$$
2\le n \le \left[\frac{2+k}{2-k}\right]
$$

with $1\le k<2$, while for $\nu_{_{-1}}=-k$ we choose $F_{1}=0,
F_{2}=0, F_{3}<0$ and so $n$ will be restricted by the inequality

$$
2\le n \le \left[\frac{3+k}{3-k}\right]
$$

with $1\le k<3$.\\

We note that for $k=1$ the possible values of $n$ are 2 and 3 for
$\nu_{_{-1}}>-1$ and $n=2$ for $\nu_{_{-1}}=-1$ i.e., naked
singularity is possible only for four and five dimensions which
we have shown in earlier works (see ref. [8-10, 20]).\\

Now we shall examine the nature of singularity by studying the
outgoing radial null geodesic (ORNG) originated from the central
shell focusing singularity. Let us start with the assumption that
it is possible to have one or more such geodesics and we choose
the form of ORNG in power series as [23, 24]

\begin{equation}
t=t_{0}+a r^{\xi},
\end{equation}

upto leading order near $r=0$ in $t$-$r$ plane with $a>0,\xi>0$ as
constants. Using equation (16) and (17) the singularity curve can
be written as (near $r=0$)
\begin{equation}
t_{s}(r)=t_{0}-\frac{F_{m}R_{0}^{\frac{n+1}{2}}}{(n+1)F_{0}^{3/2}}~r^{m}
\end{equation}

where $m\ge 1$ is an integer and $F_{m}$ is the first
non-vanishing term beyond $F_{0}$. As for naked singularity we
have $t<t_{s}(r)$ so comparing (22) with (21) for ORNG the
restrictions on the two parameters $\xi$ and $a$ as

\begin{equation}
\xi\ge m ~~~~ \text{and} ~~~~
a<-\frac{F_{m}R_{0}^{\frac{n+1}{2}}}{(n+1)F_{0}^{3/2}}
\end{equation}

Moreover, from the metric (1) we can write for ORNG

\begin{equation}
\frac{dt}{dr}=R'+R~\nu'
\end{equation}

We shall now examine the feasibility of the null geodesic
starting from the singularity with the above restrictions for the
following two cases namely, $\xi>m$ and $\xi=m$.\\

When $\xi>m$  then near $r=0$ the solution for $R$ in (9)
(choosing $t_{i}=0$) simplifies to

\begin{equation}
R=R_{0}\left(-\frac{F_{m}}{2F_{0}}\right)^{\frac{2}{n+1}}r^{\frac{2m}{n+1}+k}
\end{equation}

Now combining (21) and (25) in equation (24) we get (upto leading
order in $r$)

\begin{equation}
a~\xi~
r^{\xi-1}=R_{0}\left(\nu_{_{-1}}+k+\frac{2m}{n+1}\right)\left(-\frac{F_{m}}{2F_{0}}\right)
^{\frac{2}{n+1}}r^{\frac{2m}{n+1}+k-1}
\end{equation}

which implies

\begin{equation}
\xi=k+\frac{2m}{n+1}~~~ \text{and}~~~
a=\frac{(\nu_{_{-1}}+\xi)R_{0}}{\xi}\left(-\frac{F_{m}}
{2F_{0}}\right)^{2/(n+1)}
\end{equation}

Now if $k<m$ then $n$ and $k$ are bounded by the same
inequalities as in (20) (with $i=m$) for the formation of naked
singularity while for $k\ge m$, there will be no upper bound of
$n$. Furthermore, we note that from equation (27), $\xi>0$ and
$a>0$ as $\nu_{_{-1}}\ge -k$. Thus we have the same conclusion as
before (in case $(b)$) and it is possible to have
consistent ORNG originated from the singularity.\\

On the other hand for $\xi=m$, we have from equation (24) using
the solutions for $R$ and $\nu$ as before, it is possible to have
naked singularity if

\begin{equation}
n=\frac{m+k}{m-k}, ~~~~max(1,\frac{m}{3})\le k <m
\end{equation}

and

\begin{equation}
a=-\frac{1}{m}\left(-\frac{n+1}{2}\sqrt{F_{0}}~a-\frac{F_{m}}{2F_{0}}R_{0}^{\frac{n+1}{2}}
\right)^{\frac{1-n}{1+n}}\left[(\nu_{_{-1}}+m)\frac{F_{m}}{2F_{0}}R_{0}^{\frac{n+1}{2}}+
\frac{1}{2}(n+1)(\nu_{_{-1}}+k)\sqrt{F_{0}}~a\right]
\end{equation}
\\
Now we shall examine whether the restriction (23) for `$a$' is
consistent with the expression `$a$' in equation (29). In fact
equation (29) takes the form

\begin{equation}
2^{\frac{2}{n+1}}bm=-\left[-(n+1)b-\zeta\right]^{\frac{1-n}{1+n}}\left[(\nu_{_{-1}}+m)\zeta+
(n+1)(\nu_{_{-1}}+k)b\right]
\end{equation}

with the transformation
\begin{equation}
a=bF_{0}^{\frac{1}{n-1}},~~~~F_{m}=\frac{\zeta
F_{0}^{\frac{m}{2}+1}}{R_{0}^{\frac{n+1}{2}}}.
\end{equation}

Since equation (30) is a real valued equation of $b$, so we must
have
$$
(n+1)b+\zeta<0~,
$$
which using (31) gives us the restriction on $a$ in equation (23).
Hence the geodesic (21) will have consistent solution for `$a$'
and `$\xi$'. So the above conclusion regarding the formation of
naked singularity is justified. Further, introducing the variable
`$\phi$' by the relation

\begin{equation}
\phi=-(n+1)b-\zeta
\end{equation}

we have seen from equation (30)

\begin{equation}
4k^{n+1}\phi^{n-1}(\zeta+\phi)^{n+1}=[2k\zeta-(n-1)(\nu_{_{-1}}+k)\phi]^{n+1}
\end{equation}

with the restriction ~$0<\phi<-\zeta$ and `$a$' will satisfy the inequality in equation (23).\\

Further we observe that for $k\ge m$, $n$ can not have any
positive integral solution and hence naked singularity is not
possible. Also this case (i.e., $\xi=m$) has no analogue with our
previous result by comparing $t_{ah}(r)$ with $t_{0}$. Lastly for
any fixed $m$, $k$ can only take those values within the limits in
equation (28) which makes $n$, a positive integer($\ge 2$). It is
to be noted that if we consider next order term in the geodesic
eq. (21) then it is possible to have a family of ORNG originated
from the central singularity (see Ref. [24]).\\

Thus the formation of naked singularity strongly depends on the
nature of the initial density and also on the choice of the
parameter $k$ involved in the initial choice of the scale factor
$R$. Therefore we conclude that formation of naked singularity
strongly depends on the choice of initial data for the physical
parameters as well as for geometric quantity. For future work, it
will be interesting to study the dominance of initial data for
physical parameters over that for geometric quantities and vice
versa for formation
of naked singularity.\\

{\bf Acknowledgement:}\\

One of the authors (U.D) thanks CSIR (Govt. of India) for the
award of a Senior Research Fellowship.\\

{\bf References:}\\
\\
$[1]$  A. Ori and T. Piran, {\it Phys. Rev. Lett.} {\bf 59} 2137 (1987).\\
$[2]$  F. J. Tipler, {\it Phys. Lett. A} {\bf 64} 8 (1987).\\
$[3]$  F. C. Mena, R. Tavakol and P. S. Joshi, {\it Phys. Rev. D}
{\bf 62} 044001 (2000).\\
$[4]$  P.S. Joshi, {\it Global Aspects in Gravitation and
Cosmology} (Oxford Univ. Press, Oxford, 1993).\\
$[5]$  P. S. Joshi and I. H. Dwivedi, {\it Commun. Math. Phys.}
{\bf 166} 117 (1994); {\it Class. Quantum Grav.}
{\bf 16} 41 (1999).\\
$[6]$  D. Christodoulou, {\it Commun. Math. Phys.} {\bf 93} 171 (1984);
R. P. A. C. Newman, {\it Class. Quantum. Grav.} {\bf 3} 527 (1986).\\
$[7]$  P.S. Joshi, N. Dadhich and R. Maartens, {\it Phys. Rev. D}
{\bf 65} 101501({\it R})(2002).\\
$[8]$ A. Banerjee, U. Debnath and S. Chakraborty, {\it
gr-qc}/0211099 (2002)(accepted in {\it Int. J. Mod. Phys. D}) .\\
$[9]$ U. Debnath and S. Chakraborty, {\it gr-qc}/0211102 (2002) .\\
$[10]$ R. Goswami and P.S. Joshi, {\it gr-qc}/02112097  (2002) .\\
$[11]$ C. Barrabes, W. Israel and P. S. Letelier, {\it Phys. Lett.
A} {\bf 160} 41 (1991); M. A. Pelath, K. P. Tod and R. M. Wald,
{\it Class. Quantum Grav.} {\bf 15} 3917 (1998); F. Echeverria,
{\it Phys. Rev. D} {\bf 47} 2271 (1993); T. A. Apostolatos and K. S. Thorne, {\it Phys. Rev. D} {\bf 46} 2435 (1992).\\
$[12]$  S. L. Shapiro and S. A. Teukolsky, {\it Phys. Rev. Lett.} {\bf 66} 994 (1991); {\it Phys. Rev. D}
{\bf 45} 2006 (1992).\\
$[13]$ T. Nakamura, M. Shibata and K.I. Nakao, {\it Prog. Theor.
Phys.} {\bf 89} 821 (1993) .\\
$[14]$ T. Harada, H. Iguchi and K.I. Nakao, {\it Phys. Rev. D} {\bf 58} 041502 (1998) .\\
$[15]$  H. Iguchi, T. Harada and K.I. Nakao, {\it Prog. Theor.
Phys.} {\bf 101} 1235 (1999); {\it Prog. Theor. Phys.} {\bf 103} 53 (2000).\\
$[16]$ K. S. Thorne, 1972 in {\it Magic Without Magic}: John
Archibald Wheeler, Ed. Klauder J (San Francisco: W. H. Freeman
and Co.).\\
$[17]$  P. Szekeres, {\it Phys. Rev. D} {\bf 12} 2941 (1975).\\
$[18]$  P. S. Joshi and A. Krolak, {\it Class. Quantum Grav.} {\bf
13} 3069 (1996); S. S. Deshingkar, S. Jhingan and P. S.
Joshi, {\it Gen. Rel. Grav.} {\bf 30} 1477 (1998).\\
$[19]$  S. M. C. V. Goncalves, {\it Class. Quantum Grav.}
{\bf 18} 4517 (2001); {\it Phys. Rev. D} {\bf 65} 084045 (2002).\\
$[20]$  U. Debnath, S. Chakraborty and J. D. Barrow, {\it gr-qc}/0305075 (2003).\\
$[21]$  P. Szekeres, {\it Commun. Math. Phys.} {\bf 41} 55 (1975).\\
$[22]$  S. Chakraborty and U. Debnath , {\it gr-qc}/0304072 (2003).\\
$[23]$  S. Barve, T. P. Singh, C. Vaz and L. Witten, {\it Class.
Quantum Grav.} {\bf 16} 1727 (1999).\\
$[24]$  U. Debnath and S. Chakraborty, {\it math-ph}/0307024.\\

\end{document}